\documentclass[11pt]{article}
\usepackage[dvips]{epsfig}
\usepackage{fullpage}

\def\##1{\underline{#1}}
\def\=#1{\underline{\underline{#1}}}

\def\+#1{\underline{\bf #1}}
\def\*#1{\underline{\underline{\bf #1}}}

\def\eps{\epsilon}
\def\epso{\epsilon_{\scriptscriptstyle 0}}
\def\muo{\mu_{\scriptscriptstyle 0}}
\def\ko{k_{\scriptscriptstyle 0}}

\def\.{\mbox{ \tiny{$^\bullet$} }}

\def\le{\left(}
\def\ri{\right)}
\def\les{\left[}
\def\ris{\right]}
\def\lec{\left\{}
\def\ric{\right\}}

\def\c#1{\cite{#1}}
\def\l#1{\label{#1}}
\def\r#1{(\ref{#1})}

\begin{document}

\noindent {\bf CORRELATION LENGTH AND NEGATIVE PHASE VELOCITY IN
ISOTROPIC DIELECTRIC--MAGNETIC MATERIALS} \vskip 0.2cm

\noindent  {\bf Tom G. Mackay$^a$ and Akhlesh Lakhtakia$^b$}
\vskip 0.2cm

\noindent {\sf $^a$ School of Mathematics\\
\noindent University of Edinburgh\\
\noindent Edinburgh EH9 3JZ, United Kingdom\\
email: T.Mackay@ed.ac.uk} \vskip 0.4cm

\noindent {\sf $^b$ CATMAS~---~Computational \& Theoretical Materials Sciences Group \\
\noindent Department of Engineering Science \& Mechanics\\
\noindent 212 Earth \& Engineering Sciences Building\\
\noindent Pennsylvania State University, University Park, PA
16802--6812\\
email: akhlesh@psu.edu} \vskip 0.4cm

\noindent {\bf ABSTRACT:}  A composite material comprising
randomly distributed spherical particles of two different
isotropic dielectric--magnetic materials is homogenized using the
second--order strong--property--fluctuation theory in the
long--wavelength approximation. Whereas neither of the two
constituent materials by itself  supports planewave propagation
with negative phase velocity (NPV), the homogenized composite
material (HCM) can. The propensity of the HCM to support NPV
propagation is sensitive
  to the distributional statistics of the constituent material
  particles, as characterized by a two--point covariance function
  and its associated correlation length.
  The scope for NPV propagation diminishes
  as the correlation length increases.

\vskip 0.2cm \noindent {\bf Keywords:} {\em homogenization,
strong--property--fluctuation theory, negative refraction}

\vskip 0.4cm

\newpage

\noindent{\bf 1. INTRODUCTION}

The descriptions of electromagnetic planewave propagation
traditionally encountered in standard textbooks generally involve
positive phase velocity (PPV)~---~that is, the phase velocity
casts a positive projection onto the time--averaged Poynting
vector. On the other hand, there is growing recognition of the
importance of negative--phase--velocity (NPV) propagation, wherein
the phase velocity casts a negative projection onto the
time--averaged Poynting vector \c{LMW_CM,Rama}. Of the many exotic
phenomenons that follow as a consequence of NPV, negative
refraction has been the focus of particular attention because of
its scientific as well as technological significance \c{Pendry04}.

Manifestations of NPV are not readily observed in naturally
occurring homogeneous materials. In contrast, artificial
\emph{metamaterials} may be conceptualized~---~and in some
instances physically realized~---~which  support NPV propagation.
To date, experimental developments with NPV--supporting
homogeneous metamaterials have been limited to
  wavelengths larger than in the visible
regime, with the micromorphology
  based on elements of complicated shapes \c{SSS,NPV_expt1,NPV_expt2}.

In a recent study, we proposed
 a simple recipe for a NPV--supporting metamaterial, based on the
 homogenization of a random assembly of two different types of  spherical particles \c{ML_MOTL}.
The two types of constituent particles, type $a$ and type $b$,
are each made of an isotropic, homogeneous
  dielectric--magnetic material, the relative permittivities being denoted by
  $\eps_{a,b}$ and the relative permeabilities by $\mu_{a,b}$.
Provided that $\eps_{a,b}$ and $\mu_{a,b}$
 lie within certain parameter ranges, with the
real parts of $ \eps_{a,b} $ being negative--valued and the real
parts of $ \mu_{a,b}$ being positive--valued (or vice versa),
   the bulk
  constituent materials  do not support NPV propagation  whereas the
 corresponding homogenized composite material (HCM) does.
  The constitutive parameters of the  HCM were estimated
  using the well--established Bruggeman homogenization formalism \c{Ward}.
 The Bruggeman approach has the advantages
other recent approaches involving NPV--supporting metamaterials
  \c{Holloway,Jylha} that (i)   the constituent particles are not resonant,  and (ii)
   it is not limited to dilute composites.

The sizes of the constituent material particles play a significant
role in determining whether or not HCMs support NPV propagation
\c{Holloway,Jylha}. Using an extended Bruggeman approach
\c{LM_AEU}, we recently reported that increasing the particle size
diminishes the scope for NPV propagation in the HCM \c{LM_MOTL}. The role
 of the distributional
statistics of the constituent material particles in
NPV--supporting metamaterials has also been highlighted recently
\c{Jylha}~---~it is this topic that we explore herein.

The Bruggeman homogenization formalism takes into account the
distributional statistics of the constituent material particles
only at the lowest order, via the volume fractions of the constituent materials.
 A different
approach is provided by the strong--property--fluctuation theory
(SPFT)  in which a comprehensive description of the distributional
statistics of the constituent material particles may be
accommodated. The provenance of the SPFT lies in wave--propagation
studies for continuous random mediums \c{Ryzhov,Frisch},
 but it has more
recently gained prominence  in the homogenization of particulate
composites \c{TK81}.  Within the SPFT,  estimates of the HCM's
constitutive parameters are calculated as successive iterates
to the constitutive parameters of a homogenous comparison medium.
The iterates are expressed in terms of correlation functions
describing the spatial distribution of the constituent material
particles. In principle, correlation functions of arbitrarily high
order can be incorporated.

The SPFT is commonly  implemented at
the second--order level of approximation wherein the   distribution
statistics are stated in terms of a two--point
correlation function and its associated correlation length
\c{Stogryn}. The correlation length is taken to be large relative
to the constituent material particles, but small relative to the
electromagnetic wavelengths. The electromagnetic responses of
constituent material particles
 within
a region of linear dimensions given by the correlation length are
mutually dependent, whereas the electromagnetic responses of
constituent material particles separated by distances much greater
than the correlation length are assumed to be independent
\c{MLW00}.

   In the following sections, the
second--order SPFT is applied to investigate the role of
correlation length in predicting whether or not an isotropic
dielectric--magnetic HCM supports NPV propagation.

As regards notational matters,
$\mbox{Re} \lec z \ric$ denotes the real part and $\mbox{Im} \lec
z \ric$ the imaginary part of a complex--valued scalar $z$. The free--space wavenumber at angular
frequency $\omega$  is written as  $\ko = \omega \sqrt{\epso
\muo}$, with $\epso$ and $\muo$ being the permittivity and
permeability of free space, respectively.

\vskip 0.4cm

\noindent{\bf 2. HOMOGENIZATION}

Let us consider  the homogenization of two constituent materials described
in the previous section.
Both constituent materials are composed of
electrically small spherical particles, which
are randomly mixed together. Since  materials $a$ and $b$ are
assumed to be passive, the principle of causality~---~combined
with the implicit time--dependence $\exp(-i\omega t)$~---~imposes
the constraint \c{Chen}
\begin{equation}
\left.
\begin{array}{l}
 \mbox{Im} \, \lec \eps_{\ell} \ric > 0
\\
 \mbox{Im} \,
\lec \mu_{\ell} \ric > 0
\end{array}
\right\}, \qquad (\ell = a, b).\,\l{causality}
\end{equation}
 Let $f_\ell \in (0,1)$
($\ell = a,b$) denote the volume fraction of phase $\ell$, with
$f_a + f_b = 1$.

The HCM  is an isotropic, homogeneous, dielectric--magnetic material.
 By application  of the  SPFT,
estimates of its relative permittivity $\eps_{\scriptscriptstyle
HCM}$ and relative permeability $\mu_{\scriptscriptstyle HCM}$ are
provided by the iterative refinement of the corresponding
constitutive parameters of a homogeneous comparison medium.
Furthermore, it transpires that the relative permittivity and
relative permeability of the comparison medium are identical to
those yielded by the Bruggeman homogenization formalism
\c{TK81,MLW00}~---~namely, $\eps_{\scriptscriptstyle Br}$ and
$\mu_{\scriptscriptstyle Br}$ \c{ML_MOTL}.

At the second--order level of approximation, the SPFT estimates
emerge as \c{TK81,MLW00}
\begin{eqnarray}
\l{e1} \eps_{\scriptscriptstyle HCM} &=& \eps_{\scriptscriptstyle
Br} \lec 1 + \frac{6 k^2_{\scriptscriptstyle Br} T \les f_a \le
\displaystyle{\frac{\eps_a - \eps_{\scriptscriptstyle Br}}{\eps_a
+ 2 \eps_{\scriptscriptstyle Br}}}\ri^2 + f_b \le
\displaystyle{\frac{\eps_b - \eps_{\scriptscriptstyle Br}}{\eps_b
+ 2 \eps_{\scriptscriptstyle Br}}}\ri^2 \ris}{ 1 - 2
k^2_{\scriptscriptstyle Br} T \les f_a \le \displaystyle{
\frac{\eps_a - \eps_{\scriptscriptstyle Br}}{\eps_a + 2
\eps_{\scriptscriptstyle Br}}}\ri^2 + f_b \le \displaystyle{
\frac{\eps_b - \eps_{\scriptscriptstyle Br}}{\eps_b + 2
\eps_{\scriptscriptstyle Br}}}\ri^2 \ris} \ric ,\\
\mu_{\scriptscriptstyle HCM} &=& \mu_{\scriptscriptstyle Br} \lec
1 + \frac{6 k^2_{\scriptscriptstyle Br} T \les f_a \le
\displaystyle{\frac{\mu_a - \mu_{\scriptscriptstyle Br}}{\mu_a + 2
\mu_{\scriptscriptstyle Br}}}\ri^2 + f_b \le
\displaystyle{\frac{\mu_b - \mu_{\scriptscriptstyle Br}}{\mu_b + 2
\mu_{\scriptscriptstyle Br}}}\ri^2 \ris}{ 1 - 2
k^2_{\scriptscriptstyle Br} T \les f_a \le \displaystyle{
\frac{\mu_a - \mu_{\scriptscriptstyle Br}}{\mu_a + 2
\mu_{\scriptscriptstyle Br}}}\ri^2 + f_b \le \displaystyle{
\frac{\mu_b - \mu_{\scriptscriptstyle Br}}{\mu_b + 2
\mu_{\scriptscriptstyle Br}}} \ri^2 \ris} \ric. \l{m1}
\end{eqnarray}
Here, $k_{\scriptscriptstyle Br} =  \ko
\sqrt{\eps_{\scriptscriptstyle Br} \mu_{\scriptscriptstyle Br}}\,$
is the wavenumber in the comparison medium, with the sign of the
square root term selected to ensure that $\mbox{Im} \lec
k_{\scriptscriptstyle Br} \ric > 0$. The distributional statistics
of the constituent material particles are taken into account via
the term
\begin{equation}
T =  \int^\infty_0 \tau(R) R \, \exp \le i k_{\scriptscriptstyle
Br} R \ri \; dR,
\end{equation}
which contains  the two--point covariance function $\tau(R)$. We
choose the simple step covariance function \c{TKN82}
\begin{equation}
\tau (R) = \left\{ \begin{array}{lcr} 1, && R \leq L\\0, && R > L
\end{array},
\right. \l{cov}
\end{equation}
with correlation length $L$. Across a range of
physically reasonable covariance functions,  the SPFT estimates
of the HCM's constitutive parameters are not particularly sensitive
to the form of $\tau (R)$ \c{MLW01}.

After utilizing the long--wavelength approximation $\left|\,
k_{\scriptscriptstyle Br} L / 2 \pi \right| \ll 1$, \r{e1} and
\r{m1} may be combined with the covariance function \r{cov} to
obtain \c{MLW00}
\begin{eqnarray}
\l{e2} \eps_{\scriptscriptstyle HCM} &=& \eps_{\scriptscriptstyle
Br} \lec 1 + \le k_{\scriptscriptstyle Br} L \ri^2 \les f_a \le
\displaystyle{\frac{\eps_a - \eps_{\scriptscriptstyle Br}}{\eps_a
+ 2 \eps_{\scriptscriptstyle Br}}}\ri^2 + f_b \le
\displaystyle{\frac{\eps_b - \eps_{\scriptscriptstyle Br}}{\eps_b
+ 2 \eps_{\scriptscriptstyle Br}}}\ri^2 \ris \le 3 + i 2
k_{\scriptscriptstyle Br} L \ri \ric,
\\
\mu_{\scriptscriptstyle HCM} &=& \mu_{\scriptscriptstyle Br} \lec
1 + \le k_{\scriptscriptstyle Br} L \ri^2 \les f_a \le
\displaystyle{\frac{\mu_a - \mu_{\scriptscriptstyle Br}}{\mu_a + 2
\mu_{\scriptscriptstyle Br}}}\ri^2 + f_b \le
\displaystyle{\frac{\mu_b - \mu_{\scriptscriptstyle Br}}{\mu_b + 2
\mu_{\scriptscriptstyle Br}}}\ri^2 \ris \le 3 + i 2
k_{\scriptscriptstyle Br} L \ri \ric. \l{m2}
\end{eqnarray}

\vskip 0.4cm

 \noindent{\bf 3. NUMERICAL RESULTS}

Let us now explore how the correlation length $L$ affects the
propensity of the chosen type of HCM to support NPV propagation. We do so by
evaluating \r{e2} and \r{m2} for specific values of
$\eps_{a,b}$, $\mu_{a,b}$, $f_a$ and $L$.

\begin{figure}[!ht]
\centering \psfull \epsfig{file=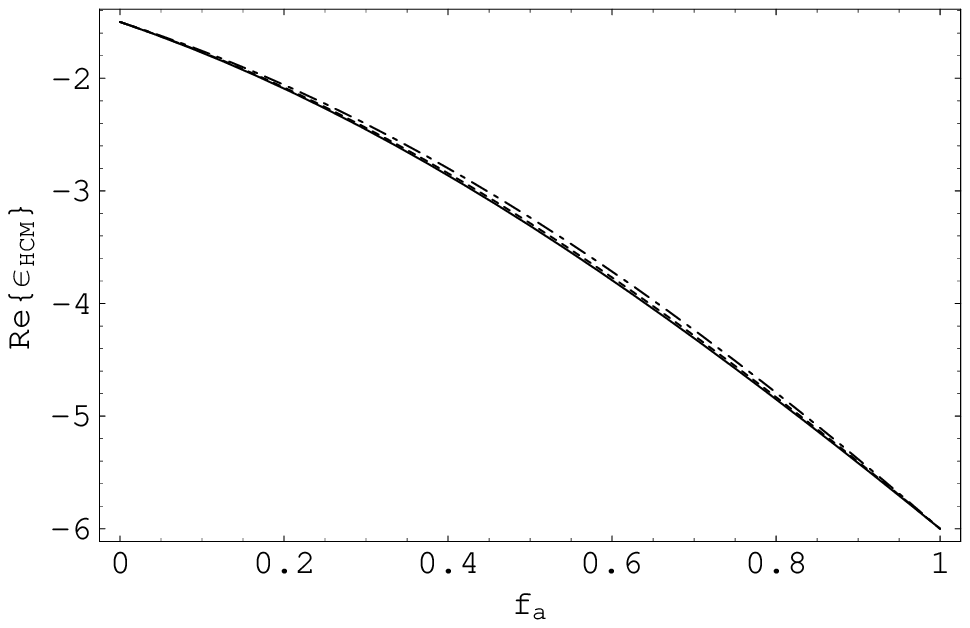,width=3.1in}
\hspace{2mm}\epsfig{file=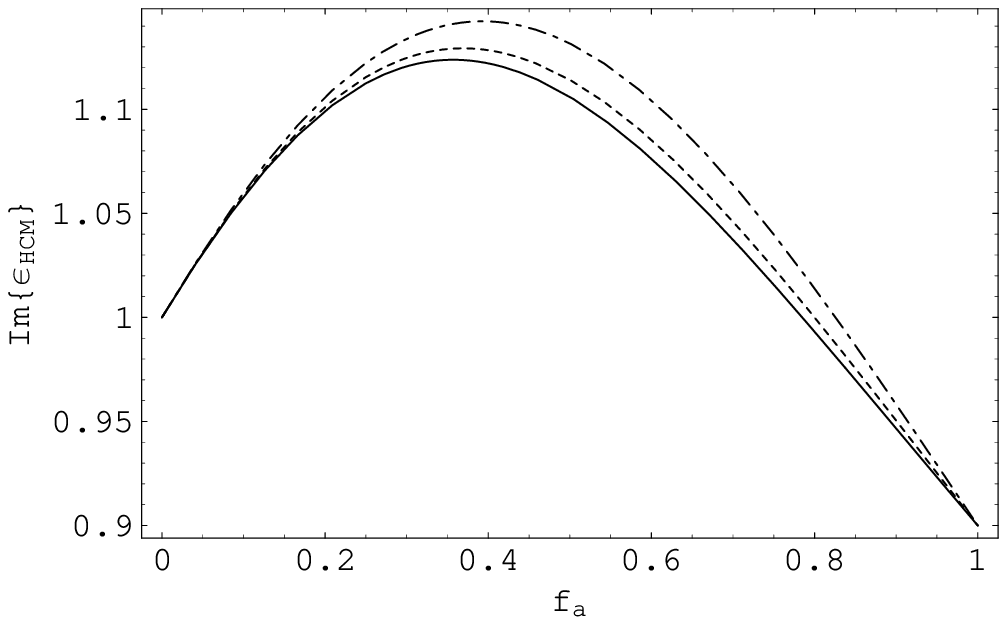,width=3.1in}\\
\epsfig{file=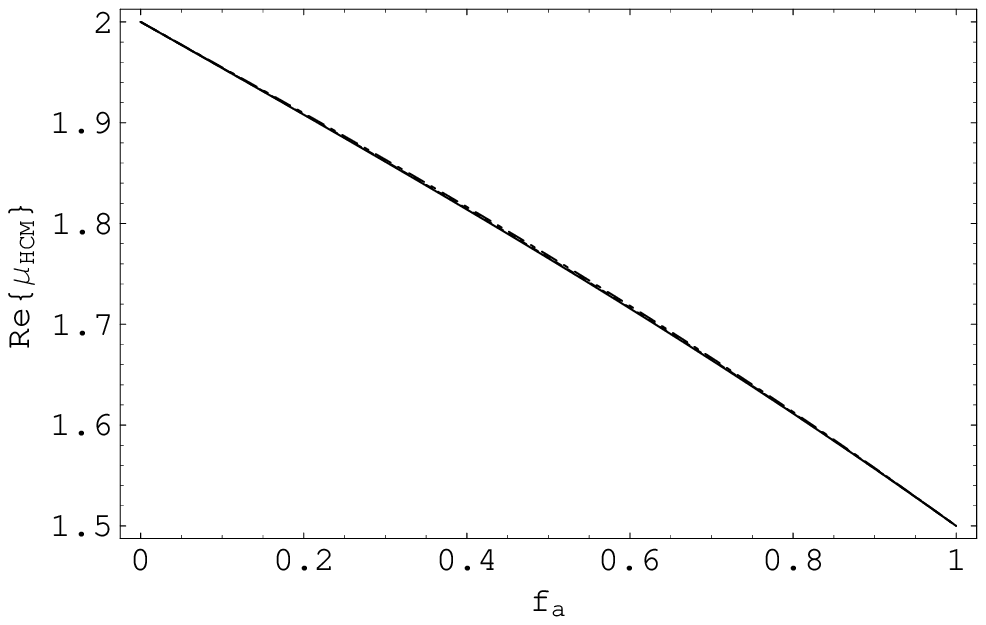,width=3.1in} \hspace{2mm}
\epsfig{file=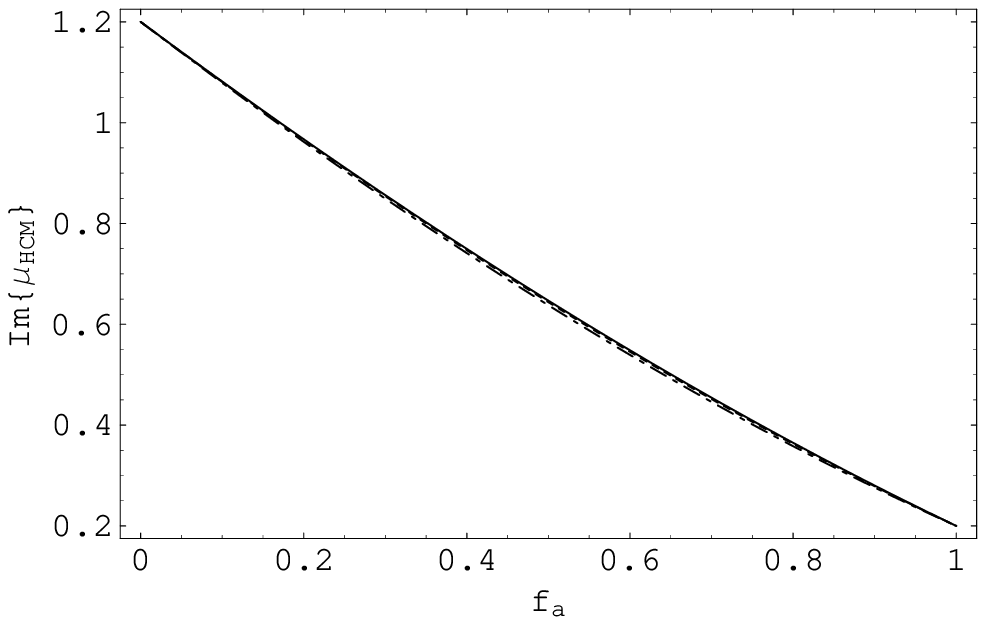,width=3.1in}
 \caption{\label{fig1} The real and imaginary  parts
 of the relative permittivity $\eps_{\scriptscriptstyle HCM}$ and the relative permeability $\mu_{\scriptscriptstyle HCM}$,
of the HCM  as
 estimated using the second--order SPFT in the long--wavelength
 approximation, plotted against volume
 fraction $f_a$ for $\ko L = 0$ (solid curve), $0.1$ (dashed curve)
and $0.2$ (broken dashed curve). Constituent material parameter
values: $\eps_a  = -6 +  0.9 i$,  $\mu_a  = 1.5 +  0.2 i$, $\eps_b
= -1.5 + i$, and $\mu_b = 2 +  1.2 i$.
  }
\end{figure}

 To allow for
direct comparison with  numerical results based on the Bruggeman
homogenization formalism reported elsewhere \c{ML_MOTL,LM_MOTL},
let us select the following values:  $\eps_a = -6 + 0.9 i $, $\mu_a  =
1.5 +  0.2 i $, $\eps_b = -1.5 + i $, and $\mu_b = 2 + 1.2 i $.
 In Figure~\ref{fig1}, the relative permittivity $\eps_{\scriptscriptstyle HCM}$
and the relative permeability $\mu_{\scriptscriptstyle HCM}$ of
the HCM are  plotted as functions of volume
 fraction $f_a$ for $\ko L \in \lec  0, 0.1, 0.2 \ric$.
  The constitutive parameters of the HCM  calculated for $\ko L
= 0$ are identical to those provided by the Bruggeman
homogenization formalism  \c{ML_MOTL}.
 The graphs are constrained such that $\mbox{Re, \,Im}
\lec \eps_{\scriptscriptstyle HCM} \ric \rightarrow \mbox{Re,\,
Im} \lec \eps_{a} \ric$ and $\mbox{Re,\,Im} \lec
\mu_{\scriptscriptstyle HCM} \ric \rightarrow \mbox{Re,\,Im} \lec
\mu_{a} \ric$ in the limit $f_a \rightarrow 1$; and similarly
$\mbox{Re,\,Im} \lec \eps_{\scriptscriptstyle HCM} \ric
\rightarrow \mbox{Re,\,Im} \lec \eps_{b} \ric$ and $\mbox{Re,\,Im}
\lec \mu_{\scriptscriptstyle HCM} \ric \rightarrow \mbox{Re,\,Im}
\lec \mu_{b} \ric$ as $f_a \rightarrow 0$.
 The correlation length influences
most obviously the imaginary part of $\eps_{\scriptscriptstyle
HCM}$: indeed,
 $\mbox{Im} \lec \eps_{\scriptscriptstyle HCM} \ric $ in Figure~\ref{fig1}
increases markedly as the correlation length increases.

\begin{figure}[!ht]
\centering \psfull \epsfig{file=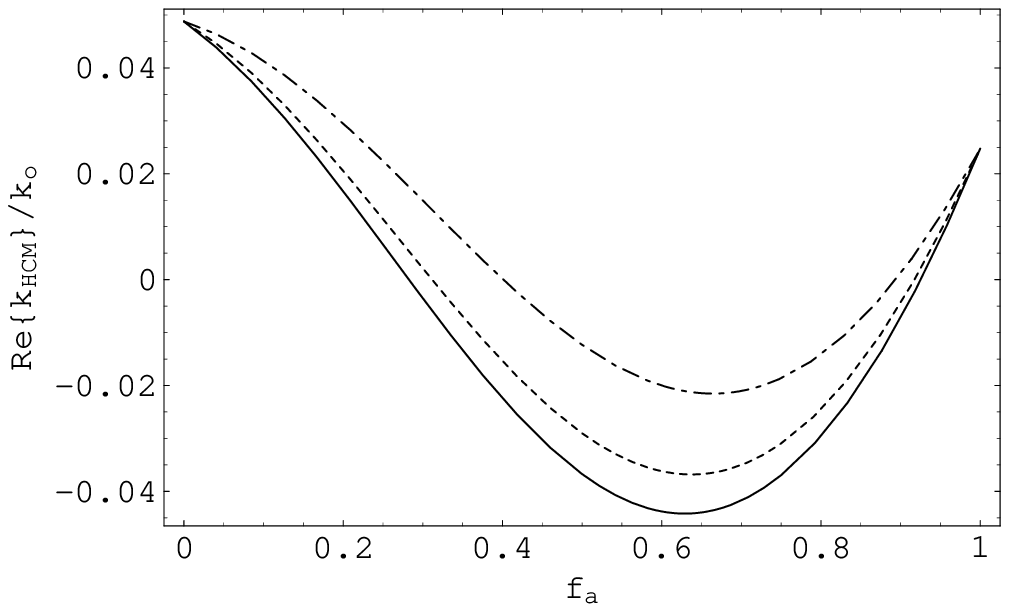,width=3.2in}\\
\hspace{10mm}\epsfig{file=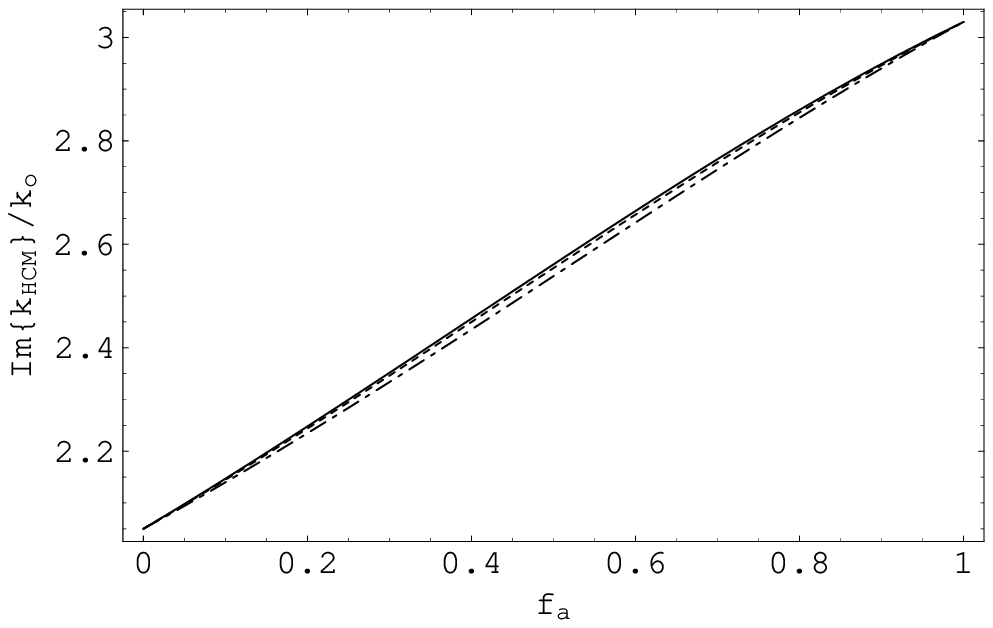,width=3.2in}
 \caption{\label{fig2}
As Figure~\ref{fig1} but for the real and imaginary  parts
 of the normalized wavenumber $k_{\scriptscriptstyle HCM} / \ko$ in the HCM.
  }
\end{figure}

The real and imaginary parts of the normalized wavenumber
\begin{equation}
\frac{k_{\scriptscriptstyle HCM}}{ \ko } =
\sqrt{\eps_{\scriptscriptstyle HCM} \mu_{\scriptscriptstyle HCM}}
\l{k_HCM}
\end{equation}
in the HCM are plotted against volume fraction $f_a$ in
Figure~\ref{fig2}, for $\ko L \in \lec  0, 0.1, 0.2 \ric$. The
sign of the square root term in \r{k_HCM} is chosen to ensure that
$\mbox{Im} \lec k_{\scriptscriptstyle HCM} \ric > 0$, as befits a
passive material. Significantly, the real part of
$k_{\scriptscriptstyle HCM}$ is negative--valued for mid--range
values of $f_a$. It is also noteworthy that $ \left| \, \mbox{Im}
\lec k_{\scriptscriptstyle HCM} \ric \, \right| \gg \left| \,
\mbox{Re} \lec k_{\scriptscriptstyle HCM} \ric \, \right| $ for
this homogenization example.

\begin{figure}[!ht]
\centering \psfull \epsfig{file=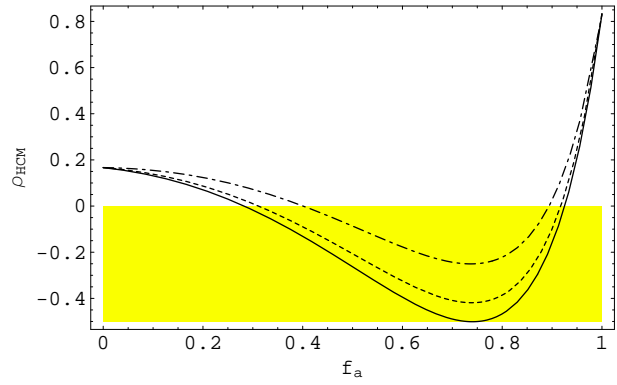,
width=3.2in}
 \caption{\label{fig3} As Figure~\ref{fig1} but for the NPV parameter  $\rho_{\scriptscriptstyle HCM}$
 in the HCM.
  Shading indicates the
region of NPV (i.e., $\rho_{\scriptscriptstyle HCM} < 0$).
  }
\end{figure}

The parameter
\begin{equation}
\rho_{\scriptscriptstyle HCM} = \frac{\mbox{Re} \, \lec
\eps_{\scriptscriptstyle HCM} \ric}{\mbox{Im} \, \lec
\eps_{\scriptscriptstyle HCM} \ric} + \frac{\mbox{Re} \, \lec
\mu_{\scriptscriptstyle HCM} \ric}{\mbox{Im} \, \lec
\mu_{\scriptscriptstyle HCM} \ric}
\end{equation}
is used to determine whether or not the HCM supports NPV
propagation \c{DL04}; NPV is indicated by
$\rho_{\scriptscriptstyle HCM} < 0$.
 In Figure~\ref{fig3}, the NPV parameter $\rho_{\scriptscriptstyle HCM}$ is graphed   against volume
 fraction $f_a$ for
 $\ko L \in \lec  0, 0.1, 0.2 \ric$.
 The positive values of $\rho_{\scriptscriptstyle HCM}$ in the limits $f_a
 \rightarrow 0$ and $f_a \rightarrow 1$ confirm that neither of
 the constituent materials $a$ and $b$ support NPV propagation.
 In contrast, the HCM clearly does support NPV propagation for
 mid--range values of $f_a$.
 The range of $f_a$ values at which the HCM supports NPV
 propagation decreases as the correlation length increases.

\begin{figure}[!ht]
\centering \psfull \epsfig{file=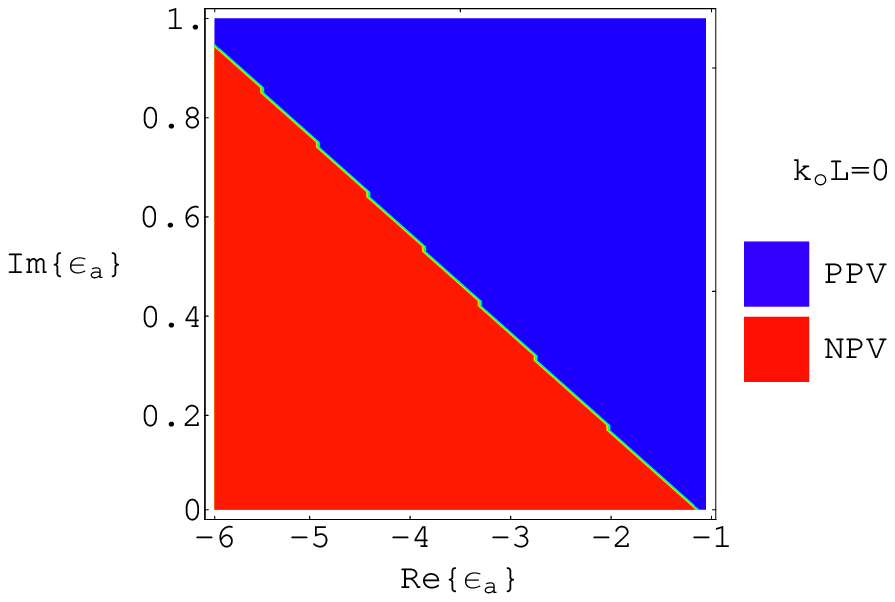, width=3.2in}
\epsfig{file=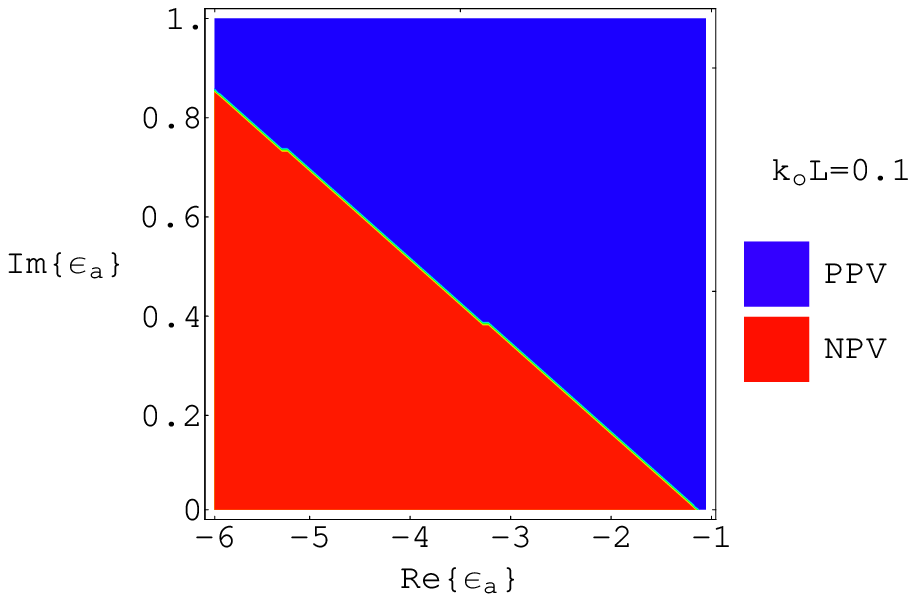, width=3.2in}\\
\epsfig{file=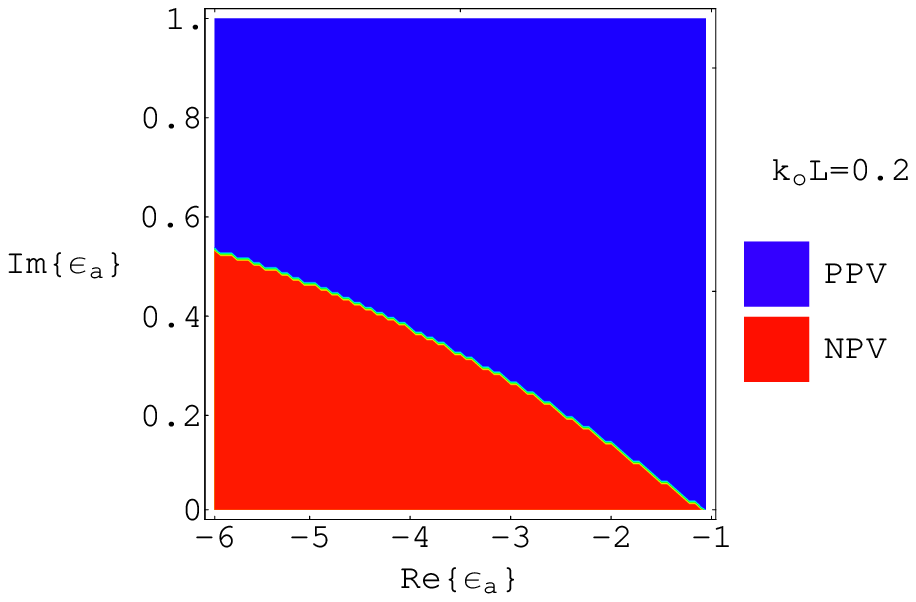, width=3.2in}
 \caption{\label{fig4} Regions of  NPV and
 PPV,
as estimated by the second--order SPFT in the
long--wavelength approximation, in relation to
 $\mbox{Re} \lec \eps_a \ric \in (-6,-1)$
and $\mbox{Im} \lec \eps_a \ric \in (0,1)$.  Whereas $\ko L  = 0$, $0.1$, and $0.2$,
the
volume fraction is fixed at $f_a = 0.3$. Other constituent material
parameter values: $\mu_a  = 1.5 + 0.2 i$, $\eps_b = -1.5 + i $, and
$\mu_b = 2 +  1.2 i$.
  }
\end{figure}

We explored this issue further in
 Figure~\ref{fig4}, wherein  regions of  NPV and
 PPV  are mapped in relation to
 $\mbox{Re} \lec \eps_a \ric \in (-6,-1)$
and $\mbox{Im} \lec \eps_a \ric \in (0,1)$. The other constituent
material parameter values are the same as those for
Figures~\ref{fig1}--\ref{fig3}; i.e., $\mu_a = 1.5 + 0.2 i $,
$\eps_b = -1.5 + i $, and $\mu_b = 2 + 1.2 i $. The volume fraction
is fixed at $f_a = 0.3$ and $\ko  L  \in \lec  0, 0.1, 0.2 \ric$.
At $\ko L = 0$ approximately half of the mapped $\eps_a$--space
supports NPV propagation, but this proportion decreases as
the correlation length increases. In particular, NPV propagation
is supported only for small values of $\mbox{Im} \lec \eps_a
\ric $ when $\ko L = 0.2$.

\vskip 0.4cm

\noindent{\bf 4. CONCLUDING REMARKS}

The isotropic dielectric--magnetic HCM arising from  a random mixture of two isotropic
dielectric--magnetic materials~---~neither of which supports NPV
propagation itself~---~supports NPV propagation within certain
parameter ranges. This conclusion, which had previously been
established by the Bruggeman \c{ML_MOTL} and the extended Bruggeman
\c{LM_MOTL} homogenization formalisms, is herein  confirmed by the
more sophisticated SPFT. In contrast to previous studies involving
NPV--supporting metamaterials \c{Holloway,Jylha}, the SPFT--based
HCM supports NPV propagation across  a wide range of volume
fraction and the prediction  that the HCM supports NPV propagation
 does not rely upon  resonant behaviour by the constituent material
particles.

By increasing the correlation length, the scope for NPV is found
to diminish, but not disappear. In this respect, the effect of
increasing the correlation length is similar to the effect of
increasing the size of the constituent material particles
\c{LM_MOTL}. That
  the correlation length and the particle size give rise to
 similar effects has
been observed elsewhere, in a different context \c{M04_WRM}.

\newpage

\newpage

\newpage

\newpage

\end{document}